\documentclass[journal,12pt,draftclsnofoot,onecolumn]{IEEEtran}

\usepackage{cite}

\usepackage{makeidx,epsfig}
\usepackage{setspace,graphicx,subfigure,multirow}

\usepackage{amsfonts,latexsym,amssymb,amsthm}

\usepackage{graphicx}
\usepackage{epstopdf}%%%%%%%%%

\usepackage{array}

\usepackage[cmex10]{amsmath}

\usepackage{url}

\usepackage{multirow}% http://ctan.org/pkg/multirow
\usepackage{hhline}% http://ctan.org/pkg/hhline

\usepackage{amsmath}
\usepackage{algorithm}
\usepackage[]{algpseudocode}
\usepackage{varwidth}

\makeatletter
\def\BState{\State\hskip-\ALG@thistlm}
\makeatother

\makeatletter
\newcommand{\algmargin}{\the\ALG@thistlm}
\makeatother
\newlength{\ifwidth}
\algdef{SE}[parIF]{parIf}{EndparIf}[1]
  {\parbox[t]{\dimexpr\linewidth-\algmargin}{%
     \hangindent\ifwidth\strut\algorithmicif\ #1\ \algorithmicthen\strut}}{\algorithmicend\ \algorithmicif}%
\algdef{C}[parIF]{parIF}{parElsIf}[1]{\parbox[t]{\dimexpr\linewidth-\algmargin}{%
     \hangindent\ifwidth\strut \algorithmicelse\ \algorithmicif\ #1\ \algorithmicthen\strut}}%
\algnewcommand{\parState}[1]{\State%
  \parbox[t]{\dimexpr\linewidth-\algmargin}{\strut #1\strut}}

\begin{document}
\bibliographystyle{IEEEtran}
\bstctlcite{IEEEexample:BSTcontrol}

\title{Efficient and Scalable Distributed Autonomous Spatial Aloha Networks via Local Leader Election}

\author{Jiangbin~Lyu,~\textit{Member,~IEEE},
        Yong~Huat~Chew,~\textit{Member,~IEEE},
        and~Wai-Choong~Wong,~\textit{Senior~Member,~IEEE}%
\thanks{Manuscript received August 05, 2014; revised February 26, 2015, July 17, 2015 and October 27, 2015; accepted January 25, 2016. The previous work \cite{PIMRC} is partially included in this paper.}%
\thanks{J. Lyu and W. C. Wong are with the Department of Electrical and Computer Engineering, National University of Singapore, Singapore 117583 (email: jiangbin.lu@u.nus.edu; elewwcl@nus.edu.sg).}%
\thanks{Y. H. Chew is with the Institute for Infocomm Research (I\textsuperscript{2}R), Agency for Science, Technology and Research, Singapore 138632 (email: chewyh@i2r.a-star.edu.sg).}%
}

\maketitle

%\vspace{-5pt}
\begin{abstract}
This paper uses a spatial Aloha model to describe a distributed autonomous wireless network in which a group of transmit-receive pairs (users) shares a
common collision channel via slotted-Aloha-like random access. The objective of this study is to develop an intelligent algorithm to be embedded into the
transceivers so that all users know how to self-tune their medium access probability (MAP) to achieve overall Pareto optimality in terms of network
throughput under spatial reuse while maintaining network stability. While the optimal solution requires each user to have complete information about the
network, our proposed algorithm
only requires users to have local information. The fundamental of our algorithm is that the users will first self-organize into a number of non-overlapping
neighborhoods, and the user with the maximum node degree in each neighborhood is elected as the local leader (LL). Each LL then adjusts its MAP according
to a parameter \textit{R} which indicates the radio intensity level in its neighboring region, whereas the remaining users in the neighborhood simply
follow the same MAP value. We show that by ensuring \textit{R} $\le 2$ at the LLs, the stability of the entire network can be assured even when each user
only has partial network information. For practical implementation, we propose each LL to use \textit{R} $=2$ as the constant reference signal to its
built-in proportional and integral controller. The settings of the control parameters are discussed and we validate through simulations that the proposed
method is able to achieve close-to-Pareto-front throughput.

\end{abstract}

\begin{IEEEkeywords}
Distributed Spectrum Sharing; Spatial Aloha; Generalized Aloha Game; Stability; Control-theoretic tuning; Pareto-front; % Local Leader Election; Distributed Spectrum Sharing
\end{IEEEkeywords}

%\vspace{-1.5pt}
\section{Introduction}

Game theory has been widely used to model the strategic interactions among intelligent devices sharing the same frequency band. In \cite{survey}, the
authors provide a comprehensive review of the game models developed for different multiple access schemes. In particular, Jin and Kesidis \cite{alohagames}
propose an Aloha game model whereby each user attempts to obtain a target rate by updating its medium access probability (MAP) in response to observed
activities. It is further assumed in \cite{alohaprice} that each user's target rate depends on its utility function and its willingness to pay, and the
authors use a pricing strategy to control these target rates to be within the feasible region. More recently, for multi-channel slotted Aloha, the joint
MAP tuning and channel selection problem is studied in \cite{CohenJSAC} under its proposed spectrum sharing model for \textit{cognitive radio}
\cite{Haykin} networks. The \textit{stability} of the strategic interactions has been studied in these works, however, they are more suitably applied to
the uplink random access channel in a fully connected centralized network.

\textit{Spatial reuse}, also known as frequency reuse, is a powerful technique to improve the area spectral efficiency of multi-user communication systems.
Cellular systems are examples whereby radios exploit the power falloff with distance and reuse the same frequency for transmission at spatially separated
locations\cite{SpatialReuseTVT}. Similar ideas can be applied in the context of \textit{dynamic spectrum access} (DSA) \cite{DynamicSpectrumAccess} and
distributed \textit{spectrum commons} \cite{brito2006spectrum} model, where different transmit-receive (Tx-Rx) pairs at a distance away are allowed to
transmit simultaneously. Chen and Huang in \cite{ChenXuSpatialReuse} study the distributed spectrum sharing problem in multi-channel slotted Aloha with
spatial reuse. The problem is formulated as a spatial channel selection game, under the assumption that the MAPs are fixed and pre-allocated. However, the
efficiency, fairness and scalability of such MAP allocations are not discussed. This paper shall address these issues for single-channel spatial Aloha
networks.

In particular, we consider a distributed wireless network in which a group of Tx-Rx pairs shares a common collision channel via slotted-Aloha-like random
access. These Tx-Rx pairs (users) are allowed to reuse the channel if they receive negligible interference from others. Such a network model is studied
using stochastic geometry by Baccelli \textit{et al.} in \cite{Baccelli09}, and named as \textit{spatial Aloha}. These approaches predict the achievable
performance of the network but give no information about network stability during its operation. Indeed, it is mathematically challenging to obtain the
stability conditions of the equilibrium solutions due to the \textit{nonsymmetric} structure in the equations formulated for a partially connected network.
How to enable the \textit{autonomous} users to self-decide and yet achieve high efficiency, good fairness and operation stability is the motivation
behind this work.

For better utilization of spatial reuse and network scalability, \textit{clustering} is used in our design. Clustering algorithms typically appear in the
context of ad hoc \cite{ClusterMANET05} and wireless sensor networks (WSN) \cite{ClusterWSN07}. These algorithms are mainly designed to perform
cluster-based routing and to achieve scalable network management, whereas our proposed clustering method tries to resolve the concurrent transmission issue
among Tx-Rx pairs, and focuses on the attainable throughput and the stability of the spatial Aloha network. The effect of clustered topologies on the
throughput of spatial Aloha is studied in \cite{ClusteredSpatialAloha} using stochastic geometry, which suggests if the nodes can adjust their transmission
parameters based on local information about their topological neighborhood, then the system performance can be improved. This also motivates our work.

Instead of viewing the distributed network from the statistical perspective, we zoom into the micro-level design of deployed networks. We specifically
develop tailored algorithms so that the users have capability to enable the network to operate at one stable equilibrium solution that is
close to the Pareto-front\cite{Pareto} throughput predicted by the \textit{generalized Aloha games} (GAG) \cite{AlohaGamesSpatialReuse}. In such games,
each Tx-Rx pair in the distributed network is a player who competes to transmit using slotted Aloha protocol. We prove the existence and uniqueness of the
Nash Equilibrium (NE) in terms of the MAP \textit{q} of all players. We further derive the stability condition of the NE. However, this condition
requires the complete knowledge on the network topology, which is neither practical nor scalable. Although we further apply a heuristic algorithm in
\cite{PIMRC} which enables the autonomous users to heuristically search for the Pareto-front throughputs based on local information obtained from
measurement, the approach nevertheless is limited by the long channel monitoring time and the poor convergence time which increase significantly with the
network size.

To provide \textit{faster convergence} to a stable operating point, we apply the control theoretic approach to tune the transmission parameters. Control
theory has been used to provide reliable and optimal configuration of 802.11 WLANs\cite{CSMAcontrol}. Similar approaches have also been applied to
implement the Distributed Opportunistic Scheduling algorithm \cite{DOScontrol}\cite{DOSgame}, however, without considering spatial reuse in their
studies. Proportional and integral (PI) controllers are adopted in the above works, and also in feedback-based clock synchronization in WSN\cite{JMchenPI}.

The novelties of our proposed \textit{Spatial Aloha via local Leader Election} (SALE) scheme are listed as follows:

$\bullet$ Each user can self-regulate its transmission parameter to ensure that the network always operates in the stable region and yet achieves
close-to-Pareto-front throughput, by using only local information about its neighbors.

$\bullet$ Rigorous stability analysis is performed. By using the stability conditions derived in \cite{AlohaGamesSpatialReuse}, we show that a local
parameter \textit{Radio Intensity Metric} (RIM), denoted as $R$, can be used to indicate the cumulative radio intensity level of each user within its
one-hop communication range --- $R \le 2$ for all users can guarantee network stability.

$\bullet$ The theory behind the design of the control system is presented. As commented by \cite{CSMAcontrol}, one of the key issues in building the control system is
to discover a \textit{constant reference signal} which relates to the desired system performance (e.g., maximum throughputs). In our case, a user which has
the maximum node degree in a certain neighborhood is elected as the \textit{local leader} (LL), and the remaining users in this neighborhood follow the
same value of MAP. Each LL adjusts its MAP according to the value of RIM computed based on its local information, and uses $R=2$ as the constant reference
signal in its built-in PI controller. As a result, the stability condition is satisfied and the MAPs are adapted towards achieving Pareto-front throughput.

$\bullet$ The PI parameters are tuned to provide fast and smooth convergence of MAP, regardless of network size.

$\bullet$ Extensive simulations are performed to verify that the SALE scheme has much faster convergence rate, better scalability and better fairness than
the heuristic algorithm, while achieving close-to-Pareto-front throughputs.

The SALE scheme may find its potential usage in DSA, where multiple Tx-Rx pairs share a common channel for transmission through spectrum
commons\cite{DynamicSpectrumAccess}, which is a spectrum resource that is owned or controlled jointly by a group of individuals\cite{brito2006spectrum}.
Imagine if these users are not equipped with proper intelligence, it may end up that the users compete among themselves for transmission since all these
selfish entities try to transmit aggressively. This could result in high contention probabilities and drive the system to function in the unstable region.
Even if all users are willing to compromise, they will not know how to achieve this objective if there is a lack of rules and regulations. This work aims to
develop an implementable methodology so that the users in a spectrum commons can work in a more controlled, more scalable, and fairer way through the
embedded intelligence.

The rest of the paper is organized as follows. We review the spatial Aloha model in Section \ref{AlohaModel}. We dedicate Section \ref{PIMRC} to give
necessary background about GAGs and the heuristic algorithm. Then we investigate the throughput optimality conditions and introduce the design parameter
RIM in Section \ref{Optimality}. Based on RIM, we design the control system for the SALE scheme in Section \ref{Local}. Then we evaluate the system
performance through simulations in Section \ref{Performance}. We conclude the paper in Section \ref{Summary}.

\section{Spatial Aloha Model}\label{AlohaModel}
Consider a distributed wireless network with $N$ transmitters, where each transmitter has its unique designated receiver. Each Tx-Rx pair is a user who
shares a common collision channel with other users, via slotted-Aloha-like random access. The conventional Aloha system is generalized to the scenario
where there exists spatial reuse among the users. If the users are located sufficiently far apart, then they can transmit in the same frequency band
simultaneously without causing any performance degradation to each other. Such a spatial reuse model can be characterized by an ``interference graph" as in
\cite{ChenXuSpatialReuse}.
The interference estimation methods in \cite{IGmessure} can be applied to obtain the interference graph.

\begin{figure}[t]
\centering
\includegraphics[width=0.35\linewidth,  trim=0 0 0 0,clip]{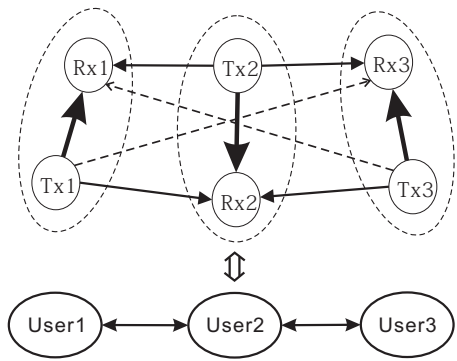} %,height=0.5\linewidth
\caption{Three Tx-Rx pairs and the interference graph\cite{AlohaGamesSpatialReuse}. In the upper part of the figure, the solid-thick arrow represents the transmission link from a transmitter to its designated receiver; the solid-thin and the dash-thin arrows represent the nonnegligible and negligible interference, respectively.} \label{pair}
%\vspace{-1.5em}
\end{figure}

As an example, three Tx-Rx pairs and the equivalent interference graph are shown in Fig. \ref{pair}, where users 1 and 3 can transmit concurrently
without collisions but neither of them can transmit together with user 2. Such interference relations can be characterized by an \textit{interference matrix}
\textbf{A}. For the topology given in Fig. \ref{pair},
\[
\textbf{A} = \begin{bmatrix}
0 & 1 & 0\\
1 & 0 & 1\\
0 & 1 & 0
\end{bmatrix}
 \]%
in which $a_{12} = 1$ means user 2 is a \textit{neighbor} of user 1, $a_{13} = 0$ means user 3 is not a neighbor of user 1, etc. We further assume that the
interference graph is an un-directed graph, then $\textbf{A}$ is a symmetric matrix, i.e., $ a_{ij}=a_{ji}, \forall i, j$. By default, $a_{ii}=0, \forall
i$.

The interference matrix characterizes the spatial distribution and frequency reuse capability of the users. Each user has different neighboring users who
directly affect its transmission. For a successful transmission for user $i$, $i\in \mathcal{N}= \{1, 2, \cdots, N\}$, all of user $i$'s neighbors (user
$j$ with $a_{ij}=1$), should not transmit. We assume that every user's transmission queue is continuously backlogged, i.e., the transmitter of every user
always has a packet to transmit to its designated receiver. Therefore, assuming that each user $i$ chooses a MAP $q_i$, then the throughput $\theta_i$ can
be obtained as:
\begin{equation}\label{throughput}
\theta_i = q_i \textstyle\prod_{a_{ij}=1}(1-q_j), \forall i \in \mathcal{N}.
\end{equation} %

\section{Close-to-Pareto-front Throughputs}\label{PIMRC}
%\vspace{-0.4em}
In this section, we summarize some of the earlier results \cite{AlohaGamesSpatialReuse}\cite{PIMRC} which form the basis of our design.

\subsection{Generalized Aloha Games}\label{generalized}
If we assume the users are selfish in nature and would choose MAP values to achieve their own objectives, then the spatial Aloha model can be formulated as
a non-cooperative game, which we name as the \textit{generalized Aloha game} (GAG) \cite{AlohaGamesSpatialReuse}. In this game, each Tx-Rx pair is a player
who competes for the channel to transmit. The objective of the game is for player $i$ to select a suitable MAP $q_i$ so that player $i$ achieves its target
rate $ y_i $, $\forall i \in \mathcal{N}$, with the lowest possible energy consumption, i.e., each player uses the smallest MAP to attain its target rate.
The target rate combination $\underline{y}=[y_1,\cdots,y_N]$ is controlled by certain pricing strategies \cite{alohaprice} or some commonly agreed
adjusting rules that try to achieve Pareto efficiency \cite{Pareto}.

We now formally state the GAG:

\textit{Players}: Distributed Tx-Rx pairs, $i\in \mathcal{N}$, who compete for a single collision channel to transmit via slotted-Aloha-like random access
scheme.

\textit{Actions}: Each player $i$ chooses a MAP $q_i\in [0,1]$, $\forall i \in \mathcal{N}$.

\textit{Objectives}: Each player $i$ ($i \in \mathcal{N}$) aims to minimize the energy consumption in attaining its target rate $y_i$, i.e.,
\begin{equation}\label{OptimizationProblem}
\begin{array}{l l}
\min ~~ q_i\\
\mbox{s.t.} ~~ \ y_i = \theta_i= q_i \prod_{a_{ij}=1}(1-q_j).
\end{array}
\end{equation}

In order to make the throughput $\theta_i$ approach the target rate $y_i$, player $i$'s myopic best response strategy in the $(m+1)$th iteration is given as:
\begin{equation}\label{iteration}
q_i^{(m+1)} = \min\{\frac{y_i}{\prod_{ a_{ij}=1}(1-q_j^{(m)})}, 1\}, \quad \forall i\in \mathcal{N}.
\end{equation}
Notice that we explicitly include the bound $q_i=1$ in (\ref{iteration}) to ensure that the mapping is within the compact domain $[0, 1]^N$. This would
introduce an extraneous solution $\underline{q}^*=\underline{1}$, which happens when the system diverges to a dead-end situation with
$\underline{q}=\underline{1}$ and $\underline{\theta}=\underline{0}$. Despite this undesirable situation, a stable NE solution would satisfy $q_i^* =
y_i/\textstyle\prod_{a_{ij}=1}(1-q_j^*), \forall i\in \mathcal{N}$. Therefore, at such an operating point $\underline{q}^*$, the throughput $\theta_i$ is
strictly equal to the target rate $y_i$, i.e., $\theta_i = q_i^*\textstyle\prod_{a_{ij}=1}(1-q_j^*)=y_i, \forall i\in \mathcal{N}$. Besides satisfying the
equality constraints in (\ref{OptimizationProblem}), we have also proved that there exists a least fixed point which enables each player to operate with
the minimal MAP concurrently. This optimal solution $\underline{q}^*$ corresponds to the unique NE of the GAG defined in (\ref{OptimizationProblem}).

The iteration process in (\ref{iteration}) is then modified as a continuous-time game to study the convergence and stability of the NE. We construct a
Lyapunov function and obtain a sufficient condition (Proposition 3 in \cite{AlohaGamesSpatialReuse}) for the stability of the NE. Specifically, the NE
$\underline{q}^*$ is stable if the matrix $\textbf{C}(\underline{q}^*)$ is positive definite, whose entries are defined as:
\begin{equation}\label{C}
\begin{array}{l l}
[\textbf{C}(\underline{q}^*)]_{ij}
=\left\{
    \begin{array}{l l}
      2 & \quad \textrm{ $i=j$}\\
      -\frac{a_{ij}q_i^*}{1-q_j^*}-\frac{a_{ji}q_j^*}{1-q_i^*} & \quad \textrm{ $i\neq j$}\\
    \end{array} \right.
\end{array}
\end{equation}

\begin{figure}[t]
\centering
\includegraphics[width=0.75\linewidth, trim=60 15 0 30,clip]{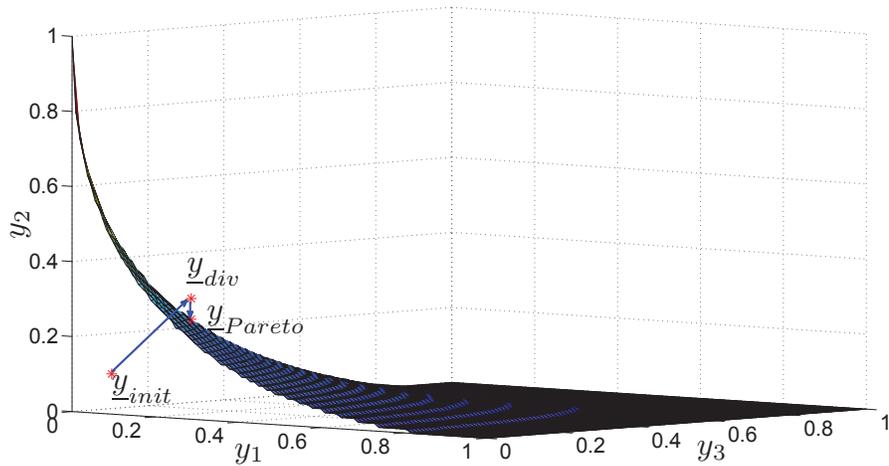}
\caption{Feasible Target Rate Region, Topology in Fig. \ref{pair} \cite{PIMRC}} \label{FeasibleRegion}
%\vspace{-1em}
\end{figure}

We now use the above condition to find a feasible region for the target rate $\underline{y}$. The word ``feasible" here means that, at the stabilized
operating point $ \underline{q}^* $, the target rate $\underline{y} $ is achievable, i.e., the throughput $ \underline{\theta}=\underline{y} $. As an
illustration, we demonstrate in Fig. \ref{FeasibleRegion} the feasible target rate region (the region under the mesh surface) for the example in Fig.
\ref{pair}. The upper boundary of this region is the Pareto front\cite{Pareto}, which is the upper bound of the feasible target rates. A network normally
operates below the Pareto front to remain stable. How to drive the users to self-regulate themselves under any topology so that the whole
network operates in the feasible region yet close to the Pareto front is the interest of this study. Note that we introduce the stability condition derived
from the GAG and use the results in our design. Thereafter, game theory is no longer the main theme, rather, the Tx-Rx pairs are
trying to set their transmission rates autonomously to achieve close-to-Pareto-front throughputs predicted by as if they are players in the GAG.

\subsection{Autonomous Pareto Optimality Achieving Algorithm}
The challenge is that, the Pareto front is determined by the network as a whole, while each user usually has only limited local information and hence does
not know how to achieve the Pareto-front throughput. As a result, if some or all the users are over-demanding, the resulting network can be unstable due to
congestions. However, if the users set low target rates, the network is stable but the channel is not fully exploited.

In \cite{PIMRC}, a set of target rate adjusting rules are introduced to enable the users to improve their throughputs
without affecting the network stability. Each user $i$ only requires local information (e.g., measured throughput $\theta_i$) to make a
myopic best response and adjust its target rate. The distributed users heuristically search for Pareto-front target rates, and the system indeed settles
down with a target rate $\underline{\hat{y}}_{Pareto}$ that is close to the Pareto front. However, the users using such a heuristic approach have to
monitor the channel activities continuously, and experience fluctuations before settling down since the network has to be driven into the
unstable region to detect the crossing of the Pareto front. As the network size increases, the system experiences more fluctuations and takes
longer to converge.
In the next few sections, we present the enabling theories behind our design of a fast self-adaptive network.

\section{Throughput Optimality Conditions}\label{Optimality}

In this section, we explain how a local parameter RIM can be defined and utilized by each user to judge for system optimality using only local information.

\subsection{Optimal Conditions}

For any MAP vector $\underline{q}=[q_1, q_2, \cdots, q_N]\in [0,1]^N$, the \textit{N} equations in (\ref{throughput}) define a vector function $\underline{\theta}(\underline{q})=[\theta_1, \theta_2, \cdots, \theta_N]$, whose value space is an \textit{N}-dimensional region. The upper boundary ($\theta_i>0, \forall i\in \mathcal{N}$) of this region is formed by the critical values of $\underline{\theta}(\underline{q})$, with the critical points $\underline{q}\in (0,1)^N$. Mathematically, this corresponds to the Jacobian matrix $\textbf{J}=\frac{\delta \underline{\theta}}{\delta \underline{q}}$ being singular\cite{maps}, i.e., the determinant of \textbf{J} at the critical point is 0:
\begin{equation}\label{Jacobian}
    \det(\textbf{J})=|\frac{\delta \underline{\theta}}{\delta \underline{q}}|=D(\underline{q})\cdot \prod\limits_{i=1}^{N}\frac{\prod_{a_{ij}=1}(1-q_j)}{1-q_i}=0\Rightarrow D(\underline{q})=0,
\end{equation}%
where
\begin{equation}\label{Dq}
D(\underline{q})=
\left|\begin{array}{cccc}
    1-q_1&-a_{12}q_1&\cdots&-a_{1N}q_1\\
    -a_{21}q_2&1-q_2&\cdots&-a_{2N}q_2\\
    \vdots&\vdots&\ddots&\vdots\\
    -a_{N1}q_N&-a_{N2}q_N&\cdots&1-q_N\\
    \end{array}\right|. \\
\end{equation}%

Similar derivations for the maximum throughputs of the original Aloha system (centralized, no spatial reuse) can be found in Section III-B of
\cite{Abramson} by Abramson. Notice that when the network is fully connected, i.e., $a_{ij}=1,\forall i\neq j$, (\ref{Jacobian}) is equivalent to formula
(26) (27) in \cite{Abramson}. In a specific case, if all MAPs are equal to $q$ in a fully connected network of $N$ users, then the maximum throughput is achieved
at $q=1/N$.

Since $D(\underline{q})$ involves all the MAPs and the complete interference matrix \textbf{A}, it is not possible for an individual user to test for this
optimal condition. In practice, generally only local information about neighbors is readily available for each user. To acquire information beyond this
will require large transmission overheads and the design will suffer from large delay. We therefore look for certain sub-optimal yet locally implementable
testing conditions. The optimal condition in (\ref{Jacobian}) gives the maximum throughput boundary and serves as a benchmark for any sub-optimal schemes.

\subsection{Sub-optimal Conditions}

From the analytical results in Section \ref{generalized}, a sufficient condition for a target rate $\underline{y}$ to be feasible is that we can find a
corresponding operating point $\underline{q}$ so that the matrix $ \textbf{C}(\underline{q}) $ is positive definite. We retrospect on the analysis in
\cite{AlohaGamesSpatialReuse} and find out a sufficient condition for $ \textbf{C}(\underline{q}) $ to be positive definite, i.e.,
$\textbf{C}(\underline{q})$ is strictly diagonally dominant\cite{matrix}:
\begin{equation}\label{LISA}
R_i=\sum\limits_{j=1,j\neq i}^{N}(\frac{a_{ij}q_{i}}{1-q_{j}}+\frac{a_{ji}q_{j}}{1-q_{i}})<2, \quad \forall i\in \mathcal{N}.
\end{equation}%a
Here we define $R_i$ as the \textit{Radio Intensity Metric} (RIM) for user $i$. In other words, if (\ref{LISA}) is satisfied, the corresponding target rate
$\underline{y}$ is achievable, i.e., at the operating point $\underline{q}$, the throughput $\theta_i$ equals to the target rate $y_i, \forall i \in
\mathcal{N}$, or the target rate falls within the feasible region (but not necessarily on the maximum throughput boundary). The converse of this, on the
other hand, is not necessarily true.

We now examine the physical meaning of RIM. Firstly, RIM is a local metric since $R_i$ consists of $q_i$ and the $q_j$ terms with $a_{ij}=1$, i.e., each
user $i$ only needs the information about its neighbors to calculate the parameter $R_i$. Secondly, from (\ref{LISA}) we observe that $R_i$ is related to
the number of user $i$'s neighbors, i.e., $N_i=\sum_{j\in \mathcal{N}}a_{ij}$. Also notice that $R_i$ is monotonic with respect to $q_i$ and $q_j
(a_{ij}=1)$. In other words, the more neighbors with the higher MAPs, the larger the value of $R_i$. Therefore, RIM can be used to indicate the cumulative
radio intensity level in the neighborhood of user $i$.

After all the $R_i, \forall i\in \mathcal{N}$ are obtained, condition (\ref{LISA}) is sub-optimal to (\ref{Jacobian}) but is now locally implementable. The
basic idea of the implementation is to tune the MAPs of all users so that condition (\ref{LISA}) is \textit{critically} satisfied, hence achieving a
sub-optimal network throughput. In other words, the MAPs of all users are tuned so that $R_l=2, \forall l\in \Omega; R_j<2, \forall j\not\in \Omega$, where
$\Omega$ is a certain subset of $\mathcal{N}$. Such an idea will be incorporated into our proposed scheme in Section \ref{Local}.

\section{The SALE Scheme}\label{Local}

In this section we present our proposed
%efficient and scalable scheme named as \textit{Spatial Aloha via local Leader Election
SALE scheme. The users first self-organize into a number of non-overlapped neighborhoods. There are many ways to approach the Pareto front surface defined
in (\ref{Jacobian}). One possible way is for the users in each neighborhood to adopt the same MAP to fulfill the fairness criterion, as it is also easier
to analyze and implement.

\subsection{Local Leader Election under Equal MAP}

From (\ref{LISA}), if we assume equal MAP in user $i$'s neighboring region, then the value of $R_i$ is related to the number of
user $i$'s neighbors given by $N_i=\sum_{j\in \mathcal{N}}a_{ij}$, which is also known as the \textit{node degree} (ND) of user $i$ in graph
theory\cite{graph}. If the MAPs of all users in a certain neighborhood are the same and gradually increase from zero, then the one with the highest ND in
this neighborhood will dissatisfy the condition (\ref{LISA}) first (assume undirected link, i.e., $ a_{ij}=a_{ji}, \forall i, j$). Therefore,
unless the users have homogeneous NDs (regular graph), their RIM values cannot reach 2 at the same time. We call the user with a locally highest RIM value
as a \textit{local leader} (LL). The key principle behind the proposed scheme is to identify these LLs, since they are most likely to cause network
instability due to interference from more neighbors. We next introduce how to identify these LLs, which consists of the following two
steps: %1) Preliminary Local Leader Election; 2) Leadership Validation.

\subsubsection{Preliminary Local Leader Election}

\begin{figure}[t]
\centering
\includegraphics[width=0.45\linewidth,  trim=0 0 0 0,clip]{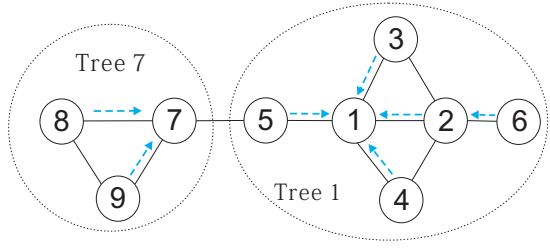} %,height=0.5\linewidth
\caption{9 Users Topology} \label{Pair9}
%\vspace{-1.5em}
\end{figure}

Only two rounds of information exchange among each user and its neighbors are needed to complete the preliminary LL election process. In the first round,
each user $i$ broadcasts its identity number (ID) $i$ to its neighbors. After the first round of broadcasting, each user $i$ will be able to compute its ND
$N_i$. In the second round, each user $i$ broadcasts its ID $i$ and its ND $N_i$, and listens from its neighbors. User $i$ then compares its ND and ID with
those of its neighbors. If $N_i$ is the largest, then user $i$ will be aware of its role as a LL (for simplicity, when two or more candidates are
connected, the one with lower ID wins). Otherwise, user $i$ would act as a \textit{follower} of its neighbors $k$ who has the highest ND (or neighbor which
has the same ND but lower ID). Such a user $k$ is called the \textit{parent} of user $i$, and user $i$ is a \textit{child} of user $k$. Note that a parent
need not be the LL. For the example in Fig. \ref{Pair9}, user 2 is the parent of user 6, but is also a child of LL 1. As a result, the whole network is
grouped into several disjoint \textit{trees}\cite{graph} (``neighborhood"), with each LL being the \textit{root} of
the tree, and the users with no children being the \textit{leaves}. The tree containing LL $l$ is thus called tree $l$, which behaves like an
independent neighborhood. The \textit{height} $H_l$ of tree $l$ is the length from the root to a leaf which is the farthest away. For the example in Fig.
\ref{Pair9}, there are two trees with LLs 1 and 7 being the roots respectively. The child-parent relationship is denoted by dashed blue arrows. Tree 7
has a height of 1, while tree 1 has a height of 2, with the longest path being $6\rightarrow 2\rightarrow 1$.

\textit{Remark}: there is no requirement for each user to have the knowledge for neighbors beyond two hops. Although a neighborhood may have
users at two or more hops away, there is no need for each node to know who is in the neighborhood. Once the child-parent relationship has been identified,
all users in the same neighborhood trace back to the same LL. The concept of a neighborhood or disjoint tree is just a virtual concept to explain the
grouping of users who move as a group while adjusting its MAP.

\subsubsection{Leadership Validation}
The essential property of a LL $l$ is that it has the locally highest RIM value $R_l$. Therefore, if $R_l\leq 2$, then all its neighbors would have a RIM
value no greater than 2, hence the stability condition in (7) is satisfied. The preliminary LL elected in the above process is the one with the highest ND.
When multiple candidates are connected, the one with the smallest ID is elected. However, choosing candidates based on smaller ID might not always
guarantee the LL to have a locally highest RIM value, in cases when the tree under consideration is affected by other trees (more details will be
discussed in Section \ref{LocalSteady}). We therefore introduce a leadership validation mechanism to handle these exceptions.

During each iteration of update, all users monitor their own RIM values. If a user $l_1$ finds that $R_{l_1}>2$, then it declares leadership and activates
its PI controller to achieve $R_{l_1}=2$. If there is a preliminary LL $l_2$ connected to user $l_1$ which hears the leadership declaration, $l_2$ should shut down its PI controller and regard user $l_1$ as its parent, i.e., both users exchange the leadership. If
there is no neighboring preliminary LL, user $l_1$ is a new LL with a separate neighborhood which consists of all its followers. In cases when two
or more connected users declare leadership, the one with smaller ID wins.

An example will be given in Section \ref{LocalSteady} to illustrate the leadership validation process.

\subsection{Control System Design}

Although the theory predicts that the network can operate in the stable region, the MAP tuning process may still exhibit oscillatory behavior if improperly
designed. An example is given by the heuristic algorithm\cite{PIMRC}, in which a user is able to detect the Pareto front solution provided that it detects
the sudden drop in its throughput when it gradually increases its MAP. By doing so, the network is already driven out of the stable region and the long
monitoring process to collect the operating parameters will significantly affect the convergence rate. To improve the tuning
process, control theoretic approach is used for each user to autonomously adapt themselves toward the sub-optimal solution.

In each update iteration, each user $i$ broadcasts its ID $i$, MAP $q_i$ and ND $N_i$ to its neighbors. After the broadcasting, each user is able to
compute its RIM value given in (\ref{LISA}). Assume that the elected LLs make up the set $\Omega$. Each LL $l\in \Omega$ sets its referenced RIM $R_{l,sp}$ to
2, and uses a PI controller\cite[Ch.10]{Astrom} to adjust its MAP $q_l$ in order to achieve $R_l=2$. Each follower $j\notin \Omega$ follows its parent $k$,
and sets $q_j(t+1)=q_k(t)$ ($q_k$ ultimately follows the MAP $q_l$ of its LL \textit{l}). When $R_l=2$ is achieved, the RIM of the followers in
tree $l$ will not be greater than 2. If $R_l=2, \forall l\in \Omega$, then the conditions in (\ref{LISA}) will be critically satisfied, thus providing a
sub-optimal network throughput. We will examine in Section \ref{DistanceToPareto} how close the design is to the Pareto front solution predicted by
(\ref{Jacobian}).
\begin{figure}[t]
\centering
\includegraphics[width=0.7\linewidth,  trim=0 0 0 0,clip]{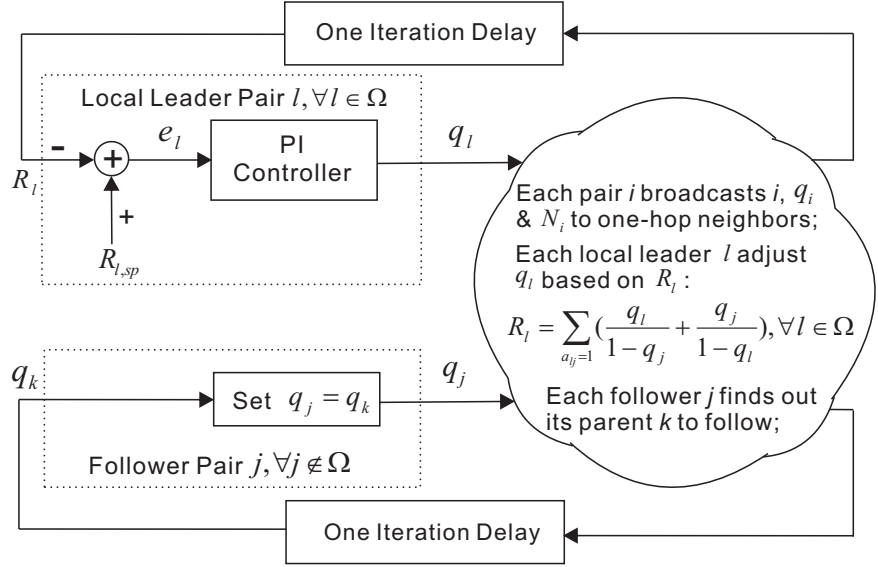} %,height=0.5\linewidth
\caption{The SALE Scheme} \label{ImplementSchemeLocal}
%\vspace{-0.5em}
\end{figure}

The control system shown in Fig. \ref{ImplementSchemeLocal} is designed as follows. For each LL $l$, the use of the PI controller is to eliminate the
steady-state error\cite[Ch. 10.1]{Astrom} while trying to achieve the desired reference signal $R_{l,sp}=2$. The relationship between the input error
signal $e_l(t)=R_{l,sp}-R_l(t)$ and the output $q_l(t)$ of the PI controller in discrete time (sampling time interval = 1) can be expressed as
%\vspace{-3pt}
\begin{equation}\label{PIcontrollerPosition}
q_l(t)=K_{P,l}e_l(t)+K_{I,l}\sigma_l(t)
\end{equation}%
where $\sigma_l(t)=\textstyle{\sum}_{n=0}^{t} e_l(n)$ is the integral function, and $\{K_{P,l}>0, K_{I,l}>0\}$ are the proportional and integral parameters
for the PI controller in LL $l$. Alternatively, we can use the recursive form below which requires no memory on the integral value $\sigma_l$:
%\cite[Ch. 3.3]{DigitalControl}
%\begin{multline}\label{PIcontroller}
%q_l(t)=q_l(t-1)+K_{P,l}[e_l(t)-e_l(t-1)]+K_{I,l} e_l(t),\\ K_{P,l}>0, K_{I,l}>0,
%\end{multline}%
\begin{equation}\label{PIcontroller}
q_l(t)=q_l(t-1)+K_{P,l}[e_l(t)-e_l(t-1)]+K_{I,l} e_l(t).
\end{equation}%

We can see from (\ref{PIcontrollerPosition}) that $q_l(t)$ stops changing after a certain time instant $t_0$, if and only if $e_l(t)=2-R_l(t)=0,\forall
t\geq t_0$ is achieved, i.e., $R_l=2$ is achieved in the steady states. Taking $z$-transform on both sides of (\ref{PIcontrollerPosition}), the transfer
function (TF) for the PI controller can be obtained as
\begin{equation}\label{PIcontrollerZ}
C_{PI}(z)=Q_l(z)/E_l(z)=K_{P,l}+\frac{K_{I,l}}{1-z^{-1}}.
\end{equation}%

The PI controller enables smooth adaptation of the MAP to achieve this desired RIM value. As commented by \cite[p.174]{Khalil}, when the goal is to
asymptotically regulate the system output (RIM) to a ``set point" ($R_{l,sp}=2$), asymptotic regulation and disturbance rejection can be achieved by
including ``integral action" in the controller. Moreover, by properly tuning the PI parameters, the PI controller can achieve a good tradeoff between the
response speed and steady-state convergence, which is a major challenge when designing adaptive algorithms.

In the next two subsections, we perform the steady-state analysis, transient analysis, and PI parameter tuning for the above control system. We first consider a simple scenario with only one LL, and then extend to multiple LLs.

\subsection{Single Local Leader Case}\label{Monotonic}

With only one LL (thus only one tree) in the network, the ND is the highest at the LL and decreases towards the leaves. For example, if tree 7 in Fig.
\ref{Pair9} does not exist and the network only comprises of tree 1, the ND decreases from the root (LL 1) towards the leaves (users 3, 4, 5, 6).

\subsubsection{Steady-State Analysis}\label{SimpleSteady}

For the above scenario, the system finally settles down with all MAPs being equal, and the RIM for the only LL $l$ at the steady state is given by:
\begin{equation}\label{SimpleLISA}
R_l=\sum\limits_{j=1,j\neq l}^{N}(\frac{a_{lj}q_{l}}{1-q_{j}}+\frac{a_{jl}q_{j}}{1-q_{l}})=\frac{2N_l \ q_l}{1-q_l}=R_{l,sp}=2.
\end{equation}%
Therefore, when operating at the steady-state, every user has a MAP equal to
\begin{equation}\label{SteadyState}
\tilde{q}_l=1/(N_l+1).
\end{equation}%
Using tree 1 in Fig. \ref{Pair9} (i.e., without tree 7) as an example, the steady-state MAPs of all users are equal to $\tilde{q}_1=0.2$ as $N_1=4$.

In particular, for any fully connected network, the LL $l$ is directly connected to all the remaining users, hence $\tilde{q}_l=1/(N_l+1)=1/N$, which
coincides with the optimal condition in (\ref{Jacobian}), i.e., Pareto-front throughput is achieved.

\subsubsection{Throughput Sensitivity on RIM}\label{SensitivitySingle}
The throughput at the LL is
\begin{equation}\label{ThroughputLeader}
\theta_l=q_l(1-q_l)^{N_l}.
\end{equation}%
When analyzing the throughput sensitivity at the LL, we examine the RIM value is perturbed by a small value $\epsilon$, i.e., $R_l=2+\epsilon$, according
to (\ref{SimpleLISA}), we have
\begin{equation}\label{qEpsilon}
q_l=(2+\epsilon)/(2+\epsilon+2N_l).
\end{equation}%
By taking the derivative of $\theta_l$ on the perturbation $\epsilon$, we can observe the local sensitivity \cite[p.251]{Boyd} of the throughput on the
value of RIM at 2.
\begin{equation}\label{ThetaToEpsilon}
\frac{\partial\theta_l}{\partial\epsilon}=\frac{\partial\theta_l}{\partial q_l}\cdot\frac{\partial q_l}{\partial\epsilon}
=\frac{2N_l\cdot[1-(N_l+1)q_l]\cdot(1-q_l)^{N_l-1}}{(2+\epsilon+2N_l)^2}.
\end{equation}%
From (\ref{ThetaToEpsilon}) we have
\begin{equation}\label{PartialEpsilon}
\frac{\partial\theta_l}{\partial\epsilon}|_{\epsilon=0}=\frac{\partial\theta_l}{\partial q_l}|_{q_l=\frac{1}{N_l+1} }\cdot\frac{\partial q_l}{\partial\epsilon}|_{\epsilon=0}=0.
\end{equation}%
Therefore, the local sensitivity of the throughput $\theta_l$ on the value of RIM at 2 is 0, which means that the throughput $\theta_l$ is locally
insensitive to small perturbations of the RIM value around 2. Similarly, we can obtain $\frac{\partial\theta_l}{\partial\epsilon}|_{\epsilon<0}>0$ and
$\frac{\partial\theta_l}{\partial\epsilon}|_{\epsilon>0}<0$. These facts suggest that the throughput $\theta_l$ is being maximized when $\epsilon=0$, i.e.,
when $R_l=2$. Moreover, when $\epsilon<0$ or equivalently $R_l<2$, there is a margin for the throughput $\theta_l$ to be improved since
$\frac{\partial\theta_l}{\partial\epsilon}>0$, i.e., user $l$ is operating at a slightly underload situation. Conversely, when $\epsilon>0$ or
equivalently $R_l>2$, the throughput $\theta_l$ decreases with $R_l$ since $\frac{\partial\theta_l}{\partial\epsilon}<0$, i.e., user $l$ is operating at a
slightly overload situation.

Now consider other users in the same neighborhood as user $l$. Firstly, if other users have the same ND as user $l$ (regular graph), then they will achieve the same MAP, RIM and throughput. Therefore, their throughputs are also maximized with a RIM value of 2. The throughput
with all users having a RIM value of 2 would be on the Pareto front. On the other hand, if other user $j$ has a smaller ND $N_j$ than that of user $l$, i.e., $N_j\leq
N_l$, then the throughput of user $j$ is
\begin{equation}\label{ThroughputFollow}
\theta_j=q_l(1-q_l)^{N_j},
\end{equation}%
where $q_l=\tilde{q}_l=1/(N_l+1)$ is the common MAP value in this neighborhood.
Since
\begin{equation}\label{PartialFollow}
\frac{\partial\theta_j}{\partial q_l}|_{q_l=\tilde{q}_l}=[1-(N_j+1)\tilde{q}_l]\cdot(1-\tilde{q}_l)^{N_j-1}=[1-(N_j+1)\frac{1}{N_l+1}]\cdot(1-\frac{1}{N_l+1})^{N_j-1}\geq 0,
\end{equation}%
hence user $j$ would be operating at the underload situation.
As a result, since the followers normally do not fully exploit the transmission opportunities, the overall throughput solution is expected to be below the Pareto front. We will address this issue again in Section \ref{DistanceToPareto}.

\subsubsection{Transient Analysis}\label{SimpleStability}

\begin{figure}[t]
\centering
\includegraphics[width=0.5\linewidth,  trim=0 0 0 0,clip]{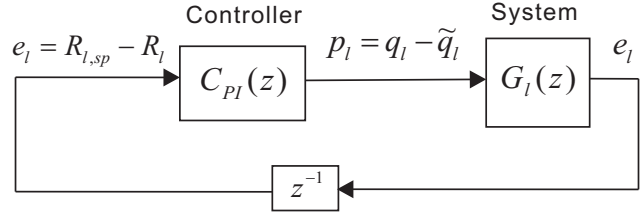} %,height=0.5\linewidth
\caption{Block Diagram for the PI Controller at the Local Leader} \label{BlockDiagram}
%\vspace{-1.5em}
\end{figure}

Before the system converges to the steady-state operating point, there exists a transient period in which $\underline{q}$ is varying. Here we use the
control theory to derive a sufficient condition to guarantee system stability. The block diagram for the PI controller at the LL is shown in Fig.
\ref{BlockDiagram}, where $C_{PI}(z)$ defined in (\ref{PIcontrollerZ}) is the TF of the PI controller, $G_l(z)$ represents the TF of the spatial Aloha
system to be controlled at the LL $l$, and $z^{-1}$ represents one sample time delay in the $z$-domain.

Define $p_l=q_l-\tilde{q}_l$ and $e_l=R_{l,sp}-R_l$ as the input and output of the system $G_l(z)$, respectively, where $\tilde{q}_l$ is the desired
operating point given in (\ref{SteadyState}), and $p_l$ is a small perturbation around $\tilde{q}_l$. For the simplicity of analysis, we would like to
simplify $G_l(z)$ by assuming no information propagation delay between the LL and its neighbors, i.e., $q_j(t)=q_l(t)$, $\forall j$ with $ a_{l,j}=1$.
The TF $G_l(z)$ can then be obtained by linearising the nonlinear function from (\ref{SimpleLISA}):
\begin{equation}\label{Sl}
R_l=(2N_l q_l)/(1-q_l )
\end{equation}%
about the operating point $\tilde{q}_l$. Eq.(\ref{Sl}) can be expressed using $p_l$ and $e_l$ as:
\begin{equation}\label{Subject}
R_{l,sp}-e_l=(2N_l (p_l+\tilde{q}_l ))/(1-(p_l+\tilde{q}_l ) ).
\end{equation}%

Taking the derivative on both sides of (\ref{Subject}), and evaluating at the operating point $p_l=q_l-\tilde{q}_l=0$, we have:
\begin{equation}\label{derivative}
\frac{de_l}{dt}=\frac{-2N_l}{[1-(p_l+\tilde{q}_l )]^2}|_{p_l=0}\cdot \frac{dp_l}{dt}=K_l\cdot \frac{dp_l}{dt},
\end{equation}%
where $K_l$ is a constant related to $N_l$:
\begin{equation}\label{K}
K_l=\frac{-2N_l}{[1-\tilde{q}_l]^2}=\frac{-2N_l}{[1-1/(N_l+1)]^2}=\frac{-2(N_l+1)^2}{N_l}.
\end{equation}%

Then we can discretize $e_l(t)$ and $p_l(t)$, and take the $z$-transform on both sides of (\ref{derivative}):
\begin{equation}\label{Ztransform}
(1-z^{-1})E_l(z)=(1-z^{-1})K_l∙P_l(z).
\end{equation}%
The TF $G_l(z)$ can then be obtained as:
\begin{equation}\label{Gz}
G_l(z)=E_l(z)/P_l(z)=K_l.
\end{equation}%

In the following we study the linearized model and ensure its stability by appropriately choosing the PI parameters. Note that the stability of the
linearized model guarantees that our system is locally stable, which means that small perturbations around the desired operating point $\tilde{q}_l$ can
all be absorbed, i.e., the control system will eventually converge to the operating point $\tilde{q}_l$ after being perturbed.

According to the control theory \cite[eq.(6.22)]{Glad}, we need to check that the following TF is stable:
\begin{equation}\label{Hz2}
H(z)=[1-z^{-1}C_{PI}(z)G_l(z)]^{-1}C_{PI}(z)\\
\end{equation}
Substituting (\ref{PIcontrollerZ}) and (\ref{Gz}), and after simplification, we have
%=[1-z^{-1}\cdot (K_{P,l}+\frac{K_{I,l}}{1-z^{-1}})\cdot K_l]^{-1}\cdot(K_{P,l}+\frac{K_{I,l}}{1-z^{-1}})\\
\begin{equation}
H(z) =\frac{(K_{P,l}+K_{I,l})z^2-K_{P,l}z}{z^2-[1+K_l(K_{P,l}+K_{I,l}) ]z+K_lK_{P,l}}.
\end{equation}
Applying the Schur-Cohn stability criterion\cite[Sec. 3.2]{AstromComputer}, a necessary and sufficient condition for a discrete-time system $H(z)$ to be
stable is that its poles all lie within the unit circle, i.e., all the roots of the characteristic equation:
\begin{equation}\label{Cz}
C(z)=z^2-[1+K_l(K_{P,l}+K_{I,l}) ]z+K_lK_{P,l}=0
\end{equation}%
should lie within the unit circle in the complex $z$-domain. Furthermore, for a second order characteristic equation $A(z)=z^2+a_1z+a_2=0$, an equivalent
stability condition is given by the Jury's stability test\cite[Theorem 3.3]{AstromComputer}:
\begin{equation}\label{Jury1}
a_2<1;\ \ \ \ a_2>-1+a_1;\ \ \ \ a_2>-1-a_1.
\end{equation}%

If we apply the Jury's stability test to (\ref{Cz}), we have:
\begin{equation}\label{triangle1}
K_lK_{P,l}<1
\end{equation}%
\begin{equation}\label{triangle2}
K_lK_{P,l}>-1-[1+K_l(K_{P,l}+K_{I,l}) ]
\end{equation}%
\begin{equation}\label{triangle3}
K_lK_{P,l}>-1+[1+K_l(K_{P,l}+K_{I,l}) ]
\end{equation}%
Since $K_l<0$, we only need $K_{P,l}>0$ to satisfy (\ref{triangle1}). Eq. (\ref{triangle3}) is equivalently reduced to $K_{I,l}>0$. From (\ref{triangle2}), we have
\begin{equation}\label{triangle4}
-K_l(2K_{P,l}+K_{I,l} )<2.
\end{equation}%
Hence a sufficient condition to guarantee stability is obtained:
\begin{equation}\label{StabilitySufficient}
\left\{
  \begin{array}{l l}
    -K_l(2K_{P,l}+K_{I,l} )<2,\\
    K_{P,l}>0, K_{I,l}>0.\\
  \end{array} \right.
\end{equation}%

\subsubsection{PI Parameter Tuning}\label{SimpleTuning}

In addition to guaranteeing stability, another consideration in selecting $\{K_{P,l}, K_{I,l}\}$ is to find a suitable trade-off between fast convergence and the transient oscillations. \textit{Ziegler-Nichols} rules\cite[Ch. 10.3]{Astrom} can be used for this purpose.

First, we compute the parameter $K_U$, which is defined as the $K_{P,l}$ value that leads to instability when $K_{I,l}=0$; and the parameter $T_I$, which is defined as the oscillation period under these conditions. According to Ziegler-Nichols rules, $K_{P,l}$ and $K_{I,l}$ can be configured as follows:
\begin{equation}\label{Ku}
K_{P,l}=0.4K_U,
\end{equation}%
\begin{equation}\label{Ti}
K_{I,l}=K_{P,l}/(0.85T_I ).
\end{equation}%

To compute $K_U$, we first set $K_{I,l}=0$ in (\ref{triangle4}), and we have
\begin{equation}\label{Kp}
K_{P,l}<1/(-K_l).
\end{equation}%
From (\ref{Kp}), we take $K_U$ as the value where the system is about to turn unstable:
\begin{equation}\label{Ku1}
K_U=1/(-K_l).
\end{equation}%
Then set $K_{P,l}$ according to (\ref{Ku}),
\begin{equation}\label{Kp1}
K_{P,l}=0.4K_U=\frac{0.4}{-K_l}=\frac{0.2N_l}{(N_l+1)^2}.
\end{equation}%

With the $K_{P,l}$ value that renders the system unstable, a given set of input values may take great changes up to every time interval, yielding an oscillation period of two time intervals ($T_I=2$). Then from (\ref{Ti}),
\begin{equation}\label{Ki}
K_{I,l}=\frac{K_{P,l}}{0.85T_I}=\frac{K_{P,l}}{1.7}=\frac{2N_l}{17(N_l+1)^2}.
\end{equation}

In summary, using the $\{K_{P,l}, K_{I,l}\}$ values given in (\ref{Kp1}) (\ref{Ki}), the SALE scheme is guaranteed to converge fast to a stable
steady-state operating point given in (\ref{SteadyState}).

\subsection{Multiple Local Leaders Case}

In order to study the scalability of the SALE scheme, we consider more general cases where there are multiple LLs in the network. We use the simple example
given in Fig. \ref{Pair9}, with users 1 and 7 being the LLs.

\subsubsection{Steady-State Analysis}\label{LocalSteady}

We have illustrated that after the LL election process, the whole network is partitioned into several disjoint trees. However, these trees
are disjoint but their MAPs are not necessarily independent of each other. In our SALE scheme, the MAPs of all users in tree $l$ are controlled by LL $l$,
who adjusts its $q_l$ based on $R_l$, and only involves its neighbors. Therefore, if LL $l$ is not directly connected to a user in other trees,
then the MAP in tree $l$ will not be affected by other trees.  For Fig. \ref{Pair9}, the MAP in tree 1 is not affected by tree 7, however, the steady-state
MAP of LL 7 is affected by user 5 which is a follower in tree 1.

The steady state of the SALE scheme can then be determined as follows. For those independent trees, the analysis in Section \ref{Monotonic} is readily
applied to obtain the steady states. For LL $l_1$ who is affected by user $m$ in tree $l_2$ (which happens normally when $N_{l_1}\leq N_{l_2}$), it should
wait until the steady state of $\tilde{q}_m=\tilde{q}_{l_2}$ is calculated before calculating $\tilde{q}_{l_1}$ based on $R_{l_1}=2$. It can be shown that to achieve $R_{l_1}=R_{l_2}=2$ with $N_{l_1}\leq N_{l_2}$, LL $l_1$ should have $\tilde{q}_{l_1}\geq 1/(N_{l_1}+1)\geq \tilde{q}_{l_2}$ in
general.
%Proceed until the steady states of all trees are calculated.
For Fig. \ref{Pair9} where $l_1=7, l_2=1, m=5$, the steady state MAP in tree 1 can be obtained from (\ref{SteadyState}) as $\tilde{q}_i=1/(N_1+1)=0.2, \forall i\in \{
1,\cdots, 6\}$. Then for LL 7, $R_7=\sum\limits_{j=5,8,9}(\frac{q_{7}}{1-q_{j}}+\frac{q_{j}}{1-q_{7}})
=4\cdot\frac{q_{7}}{1-q_{7}}+\frac{q_{7}}{1-\tilde{q}_{5}}+\frac{\tilde{q}_{5}}{1-q_{7}}$. By setting $R_7=2$ and substituting in $\tilde{q}_5=0.2$, we
have $\tilde{q}_7=0.2598$. Thus the steady state MAP in tree 7 is $\tilde{q}_i=0.2598, \forall i\in \{ 7,8,9\}$. It is easily verified that
$\tilde{q}_7>1/(N_7+1)=0.25>\tilde{q}_1$. Notice that the steady state MAP in tree 7 is \textit{not} trivially equal to $1/(N_7+1)$ as in the single LL
case. Fortunately, our control system is able to converge to the steady state \textit{automatically}.

\begin{figure}[t]
\centering
\includegraphics[width=0.5\linewidth,  trim=0 0 0 0,clip]{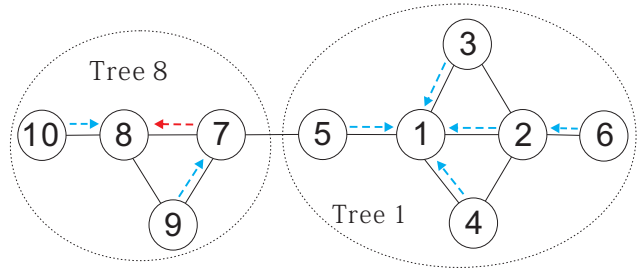} %,height=0.5\linewidth
\caption{10 Users Topology} \label{Pair10_8}
%\vspace{-1.5em}
\end{figure}

Now we illustrate the leadership validation process by adding another user 10 as a neighbor of user 8, as shown in Fig. \ref{Pair10_8}. Here user 1 and user 7 are the preliminary LLs. User 8 now
has the same ND as user 7, but user 7 is still elected as the preliminary LL due to its smaller ID. Using the above analysis, the preliminary LL 7
would push the MAP in the tree to be $\tilde{q}_i=0.2598, \forall i\in \{ 7,8,9,10\}$, which achieves $R_7=2$ but this would push the RIM of user 8 to
$R_8=2.1>2$, which should be avoided. Somewhere in the process of following the MAP of user 7, user 8 should find itself a more
suitable LL than user 7. This example shows that the leadership should not be finalized simply based on smaller ID. In such a case, the leadership should
be handed over to user 8. For the above example, in the midst of updating, when user 8 detects $R_8>2$, it declares leadership and activates its PI
controller to achieve $R_8=2$. The preliminary LL 7 becomes a follower whose parent is the new LL 8. For simplicity, the child-parent relationship is only
adjusted between the new and old LLs, while other followers stick to their original parent. For example, follower 9's parent is still user 7. The new
child-parent relationship is shown in Fig. \ref{Pair10_8}. Notice that after the leadership handover, the new LL 8 is not connected to any users in tree 1,
hence the tuning of MAP in tree 8 becomes independent of tree 1. The equilibrium MAP in the tree led by LL 8 is 0.25, while $R_8=2$ and $R_7=1.91$.

\subsubsection{Transient Analysis and PI Parameter Tuning}\label{LocalStability}

The stability is guaranteed through the following argument. When all trees are independent, the transient analysis and PI parameter tuning in each tree
follow those in Section \ref{Monotonic}, hence stability is guaranteed. Each LL $l$ has its own PI controller, with the parameters $K_{P,l},K_{I,l}$ given
in (\ref{Kp1}) and (\ref{Ki}), respectively. Again the PI parameters only rely on the local information $N_l$, hence can be immediately set after the LL
election. The system TF to be controlled at LL $l_1$ follows (\ref{Gz}) after applying (\ref{K}):
\begin{equation}\label{Gl1}
G_{l_1}(z)=K_{l_1}=-2(N_{l_1}+1)^2/N_{l_1}.
\end{equation}%

If LL $l_1$ is affected by user $m$ in tree $l_2$ (e.g., see Fig. \ref{Pair9}), this happens normally when $N_{l_1}\leq N_{l_2}$. In this case, tree $l_2$
is independent of tree $l_1$, but tree $l_1$ depends on tree $l_2$ through user $m$. Tree $l_2$ would reach the steady state first, with
$\tilde{q}_m=\tilde{q}_{l_2}\leq 1/(N_{l_1}+1)$, which remains constant afterwards. For tree $l_1$, due to the impact of $\tilde{q}_m$, from (\ref{Sl}) the
RIM for LL $l_1$ becomes:
\begin{equation}\label{Rl1}
R'_{l_1}=\sum\limits_{j=1,j\neq l_1}^{N}(\frac{a_{l_1j}q'_{l_1}}{1-q_{j}}+\frac{a_{jl_1}q_{j}}{1-q'_{l_1}})=\frac{2(N_{l_1}-1)q'_{l_1}}{1-q'_{l_1}}+\frac{q'_{l_1}}{1-\tilde{q}_m}+\frac{\tilde{q}_m}{1-q'_{l_1}},
\end{equation}%
where $\tilde{q}_m$ is now a constant, i.e., $d\tilde{q}_m/dt=0$.
In (\ref{Rl1}), we have used $'$ to denote the respective parameters to differentiate from the independent case.
As a result of \textit{fewer neighboring followers}, $R'_{l_1}$ reacts more slowly to $q'_{l_1}$ than in the independent case, hence we expect the absolute gain $|K_{l_1}|$ of the system $G_{l_1}(z)$ to decrease.

Specifically, if we apply the same linearization procedure in Section \ref{SimpleStability} to (\ref{Rl1}), then the system $G'_{l_1}(z)$ to be controlled
at the LL $l_1$ is:
\begin{equation}\label{Gl1prime}
G'_{l_1}(z)=K'_{l_1}=-\frac{2(N_{l_1}-1)}{(1-\tilde{q}'_{l_1})^2}-\frac{1}{1-\tilde{q}_m}-\frac{\tilde{q}_m}{(1-\tilde{q}'_{l_1})^2},
\end{equation}%
where $\tilde{q}'_{l_1}$ is the operating point that achieves $R'_{l_1}=2$ in (\ref{Rl1}). It can be verified using Mathematica\cite{mathematica} that
$|K'_{l_1}|<|K_{l_1}|$ for all $N_{l_1}\geq 2$ and $0\leq \tilde{q}_m\leq 1/(N_{l_1}+1)$. Similar results can be verified if LL $l_1$ is affected by more
neighbors $m_1, m_2,\cdots$ that belong to other trees.

Then the PI parameters $\{K'_{P,l_1},K'_{I,l_1}\}$ that guarantee system stability can be set from (\ref{Kp1}) and (\ref{Ki}):
\begin{equation}\label{Kp1prime}
K'_{P,l_1}=\frac{0.4}{-K'_{l_1}}>\frac{0.4}{-K_{l_1}}=K_{P,l_1}=\frac{0.2N_{l_1}}{(N_{l_1}+1)^2},
\end{equation}%
\begin{equation}\label{Kiprime}
K'_{I,l_1}=\frac{0.4}{-1.7K'_{l_1}}>\frac{0.4}{-1.7K_{l_1}}=K_{I,l_1}=\frac{2N_{l_1}}{17(N_{l_1}+1)^2}.
\end{equation}
Since $\{K_{P,l_1},K_{I,l_1}\}$ are easy to obtain using $N_{l_1}$ only, LL $l_1$ uses $\{K_{P,l_1},K_{I,l_1}\}$ in practice. More importantly, since the
current system $G'_{l_1}(z)$ has a smaller absolute gain $|K'_{l_1}|$ due to fewer neighboring followers, LL $l_1$ is using the less aggressive PI
parameters $\{K_{P,l_1},K_{I,l_1}\}$ (see Section \ref{SimpleTuning}) and hence system stability is guaranteed.

\subsubsection{Throughput Sensitivity on RIM}\label{SensitivityMultiple}
Following the above settings, assume that the LL $l_1$ is affected by user $m$ in tree $l_2$ whose steady state MAP is $\tilde{q}_m$, its throughput is given by
\begin{equation}\label{ThroughputLeaderL1}
\theta_{l_1}'=q'_{l_1}(1-q'_{l_1})^{N_{l_1}-1}(1-\tilde{q}_m).
\end{equation}%
When the RIM value in (\ref{Rl1}) is perturbed by a small value $\epsilon$ around 2, i.e., $R'_{l_1}=2+\epsilon$, by taking derivative on both sides of (\ref{Rl1}), we have $\frac{\partial q'_{l_1}}{\partial\epsilon}=-\frac{1}{K'_{l_1}}>0$.
Notice that when $\epsilon=0$, i.e., $R'_{l_1}=2$, it can be verified from (\ref{Rl1}) that the operating point $\tilde{q}'_{l_1}$ should be no greater than $1/N_{l_1}$.
Therefore, the local sensitivity of throughput $\theta_{l_1}'$ on RIM $R'_{l_1}$ around 2 is
\begin{equation}\label{PartialEpsilonL1Theta}
\frac{\partial\theta_{l_1}'}{\partial\epsilon}|_{\epsilon=0}=\frac{\partial\theta_{l_1}'}{\partial q'_{l_1}}|_{q'_{l_1}\leq\frac{1}{N_{l_1}} }\cdot\frac{\partial q'_{l_1}}{\partial\epsilon}|_{\epsilon=0} \geq 0
%(1-N_{l_1}\cdot\frac{1}{N_{l_1}})\cdot(1-\frac{1}{N_{l_1}})^{N_{l_1}-2}\cdot(1-\tilde{q}_m)\cdot(-\frac{1}{K'_{l_1}})= 0.
\end{equation}%
Therefore, in such cases there are still some margins for the throughput $\theta_{l_1}'$ to be improved, i.e., the network around the location of LL $l_1$
is operating at a slightly underload situation. Similar conclusions can be drawn when the LL $l_1$ is affected by more users in other trees. We next discuss how close the throughput obtained by SALE is to the Pareto front.

\subsection{``Distance" to Pareto Front}\label{DistanceToPareto}

The Pareto front surface is obtained if we apply the sufficient and necessary testing criteria (\ref{Jacobian}) to the network. Any point on this surface
gives a combination of throughputs achievable by all users while keeping the network operating in a stable condition. The solution obtained in
SALE generally stays below the Pareto front due to two reasons. Firstly, the stability criteria (\ref{LISA}) used in implementing the algorithm is only a
sufficient condition. Secondly, some of the followers may not have fully exploited the transmission
opportunities. Hence, the feasible throughput region obtained by SALE is only a subset to that obtained by using (\ref{Jacobian}).

Based on the sensitivity analysis in Section \ref{SensitivitySingle} and Section \ref{SensitivityMultiple}, in the homogeneous ND case (regular graph) all users have a RIM value of 2, hence the throughput solution stays
on the Pareto front. However, in most practical cases where there are variations in the NDs of users, the LL election allows partitioning the network into
several local neighborhoods. Each LL, which has the highest ND in its neighborhood, uses a PI controller to achieve a RIM value of 2. The remaining nodes
in the same neighborhood have a smaller ND than its LL, and will have a RIM value no greater than 2. This suggests that normally the followers are
operating at the underload condition, or at a distance below the Pareto front obtained by using (\ref{Jacobian}).

We attempt to characterize such a throughput margin with the optimal one obtained in (\ref{Jacobian}) by defining the ``distance to Pareto" $d_{pareto}$.
When we obtain a solution $\underline{\theta}=[\theta_1,\theta_2,\cdots,\theta_N]$ in the SALE scheme, we continue to move in the direction
$d\cdot\underline{\theta}$ ($d\geq 1$, i.e., proportionally increase the throughput of all users) until we find an operating point $\underline{q}$ that
achieves $d_{pareto}\cdot\underline{\theta}$, and beyond this point there is no stable solution. In particular, when $d_{pareto}=1$, the solution is on the
Pareto front. In the simulation results, the SALE scheme achieves a close-to-Pareto-front throughput, with $d_{Pareto}$ below 1.05 for most of the
topologies, i.e., less than 5\% below the Pareto front.

\subsection{Complexity, Scalability and Overhead of SALE}\label{SectionComplexity}

\begin{algorithm}
\caption{The SALE Scheme}\label{SALEpseudo}
\begin{algorithmic}[1]
\small \BState \textbf{Preliminary Local Leader Election}: \State Each user $i$ broadcasts its ID to its neighbors; \State Each user $i$ computes its ND
$N_i$; \State Each user $i$ broadcasts its ID and $N_i$ to its neighbors; \State Each user $i$ compares $N_i$ with the NDs of its neighbors. If $N_i$ is
the largest, then user $i$ is elected as a LL. Otherwise, user $i$ is a follower, whose parent is the neighbor with the largest ND. In cases when two or
more candidates are connected, the one with smaller ID wins. Assume the LLs make up a set $\Omega\subset \mathcal{N}$.

\ \BState \textbf{Control System}: 
\State  
    $R_{l,sp}=2$, $K_{P,l}=0.2N_l/(N_l+1)^2, K_{I,l}=2N_l/[17(N_l+1)^2], \forall l\in\Omega$;
\State $t=0$,
$\underline{q}(t)=\underline{0}$; $Declare_i=0,\forall i\in \mathcal{N}$; 
\Repeat: 
\State Each user $i$ broadcasts its MAP $q_i(t)$ to its neighbors;
\State 
\begin{varwidth}[t]{0.9\linewidth}
Each user $i$ computes $R_i(t)=\textstyle\sum_{j=1,j\neq i}^{N}[a_{ij}q_{i}(t)/(1-q_{j}(t))+a_{ji}q_{j}(t)/(1-q_{i}(t))]$;
\end{varwidth}
\For {each LL
$l\in \Omega$}: 
\State $e_l(t+1)=R_{l,sp}-R_l(t)$; 
\State $q_l(t+1)=q_l(t)+K_{P,l}[e_l(t+1)-e_l(t)]+K_{I,l} e_l(t+1)$; \EndFor 
\For {each follower
$j\not\in \Omega$}: 
\State $q_j(t+1)=q_k(t)$, where user $k$ is the parent of user $j$; 
\EndFor 

\ \State \textbf{Leadership Validation}: 
\For {each
user $i\in\mathcal{N}$}: 
\parIf {$Declare_i=1$}
\parState {
User $i$ joins $\Omega$ and becomes a LL. In cases when two or more connected users declare
leadership, the one with smaller ID wins. 
}
\parElsIf {
user $i\in\Omega$ and its neighbor declares leadership
}
\parState {
the preliminary LL quits from $\Omega$,
and follows the newly declared LL; 
}
\EndparIf 
\If {$R_i(t)>2$}: 
\State 
\begin{varwidth}[t]{0.8\linewidth}
mark $Declare_i=1$ and declare leadership in the next round of broadcast; 
\end{varwidth}
\Else
$Declare_i=0$; 
\EndIf 
\EndFor 

\ \State $t=t+1$; 
\Until{Convergence, i.e., $R_l(t)=R_{l,sp},\forall l\in\Omega$.}
\end{algorithmic}
\end{algorithm}

We summarize the SALE scheme in Algorithm 1. The proposed scheme shows the following advantages:
\subsubsection{Low Implementation Complexity}
\textit{a)} It takes only two rounds of information exchange among each user and its neighbors to complete the preliminary LL election. \textit{b)} In each
iteration, each user only needs to broadcast its ID, MAP, and ND to its neighbors. \textit{c)} Each user only uses information about its neighbors to
update its MAP in each iteration. \textit{d)} Each LL $l$ implements a simple PI controller to adjust its MAP $q_l$ so as to achieve $R_l=2$, which
corresponds to a throughput close to the Pareto front. \textit{e)} The PI parameter tuning can be autonomously done by the LL alone based on its ND $N_l$,
which guarantees stability and fast convergence. \textit{f)} Each follower $j$ only needs to find out its parent $k$, and simply sets $q_j(t+1)=q_k(t)$ in
each iteration. \textit{g)} Only in some situations, there is a need to change the LL. Therefore, SALE can be implemented autonomously with low
complexity.

\subsubsection{High Scalability}
\textit{a)} SALE is fully autonomous without any centralized controller. \textit{b)} The whole network is grouped into several disjoint trees,
and there is a LL controlling the MAPs in each tree. \textit{c)} The Ziegler-Nichols rules adapt the PI parameters in (\ref{Kp1}) (\ref{Ki})
to various \textit{user densities} (UDs) (associated with different ND $N_l$ at the LL), thus guaranteeing fast and smooth convergence of MAPs at the LLs. \textit{d)}
Given a certain UD, if the number of users increases, the number of LLs also increases correspondingly, i.e., the whole network is grouped into
more trees. As a result, the average number of users in a tree, as well as the tree height, will not change significantly as the network size grows.
Therefore, when a LL $l$ reaches the steady state ($R_l=2$), the corresponding MAP $q_l$ would not take too many hops to reach all the followers in tree
$l$. \textit{e)} As we will see in Section \ref{SimulationScalability}, the SALE scheme converges in around 40 iterations, regardless of the UD
or the number of users in the network. Therefore, SALE provides fast convergence with high scalability.

\subsubsection{Overhead of Information Exchange}
SALE requires local information exchange for leader election and MAP adaptation. The LL election requires two rounds of local information
exchange about node ID and ND. Thereafter, the MAP adaptation requires each user to broadcast its ID and MAP to its one-hop neighbors.

As mentioned in \cite{AlohaOverhead} to handle the case with information exchange, slotted Aloha usually has a framed structure consisting of a control
phase for the information exchange and a normal phase for data transmission.
%The overhead of information exchange can be calculated as follows.
In our case, we embed the message in the packet header. In a simplified model, assume that each packet originally (i.e. in the heuristic approach) has a
header field and constant packet size, which at least contains the user ID or address. For the need of our algorithm, we add three subfields to the header,
i.e., ND subfield, MAP subfield and leadership declaration subfield, and each occupies 8 bits, 16 bits and 1 bit, respectively. We further assume each
packet occupies a time slot and has a packet size of $L_S$ bits, e.g., 250 bytes = 2000 bits, the newly added fractional overhead is then given by
$25/L_S=25/2000=0.0125$.

The message exchange is realized in the following way. In each time slot, each user either sends a message to its neighbors according to its MAP, or
listens to the channel to receive a message from its neighbors. Assume that each iteration in the SALE scheme corresponds to packet transmissions in $L_F$
slots which is known as a frame, and all users are frame-synchronized. Since the transmission of a packet is subject to collision, some packets (and hence
the added subfields) will not be received correctly by some users. However, for a sufficiently large $L_F$ value, each user is likely to receive at least
one packet from all its neighbors and hence gather enough information about all its neighbors through the subfields embedded in the received packets. By
the end of each iteration, all users then update the MAP subfields for the next iteration according to the SALE scheme. Since the SALE scheme relies on an
accurate estimation of ND, we assume that each user counts and updates its ND in every $L_{ND}$ slots. $L_{ND}$ can be chosen to be sufficiently large to
guarantee an accurate estimation of ND. In our simulations, we choose $L_{ND}=10L_F$ slots (i.e., 10 iterations) to guarantee an accurate estimation.

\section{Performance Evaluation}\label{Performance}

In the simulations we set the channel bit rate at $r_b=20$ Mbps. Each packet in slotted Aloha has $L_S=2000$ bits. The slot time is
$L_S/r_b=0.1$ ms. Each iteration in SALE corresponds to a frame of $L_F=100$ slots (i.e., 10 ms), and the users are frame-synchronized. Each user counts
and updates its ND in every $L_{ND}=10L_F=1000$ slots (i.e., 10 iterations).

\begin{figure}[t]
\centering
\includegraphics[width=0.65\linewidth,  trim=0 0 0 0,clip]{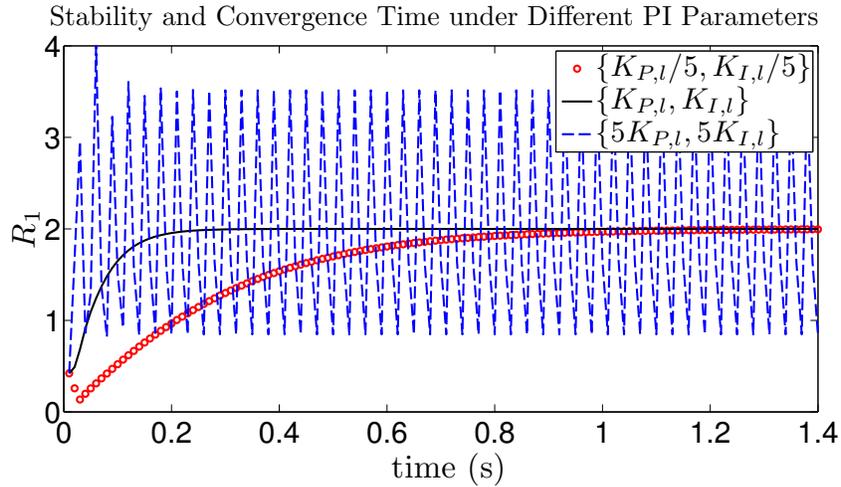} %,height=0.5\linewidth
%\vspace{-0.5em}
\caption{Stability and Convergence Time under different PI parameters} \label{PItuneCompare}
%\vspace{-1.5em}
\end{figure}

\subsection{Parameter Tuning: Stability and Convergence Time}
In Section \ref{Local} we claim that the PI parameters given in (\ref{Kp1}) (\ref{Ki}) are able to guarantee system stability as well as fast convergence
to the steady-state operating point. Here we illustrate this by using the example in Fig. \ref{Pair10_8}. Three sets of PI parameters are used in the PI
controllers respectively: $\{K_{P,l}/5,K_{I,l}/5\}$, $\{K_{P,l},K_{I,l}\}$, and $\{5K_{P,l},5K_{I,l}\}$, where $\{K_{P,l},K_{I,l}\}$ are obtained by
(\ref{Kp1}) (\ref{Ki}). The algorithm starts with small initial MAPs, e.g., $q_i=0.05, \forall i$. The transient behaviors of $R_1$ are plotted in Fig.
\ref{PItuneCompare}. Each iteration in SALE transmits $L_F=100$ packets. The value of $L_F$ is arbitrary and with the purpose to safe-guard the correct
reception of neighbors' information. This is similar to the use of $L_{ND}$. In fact, instances of message passing failure in a frame are
rarely captured in our simulations. In the rare occasion if it happens, the transient variations caused by the delay of MAP feedback are well handled by
the control system.

From Fig. \ref{PItuneCompare} we can see that $\{K_{P,l},K_{I,l}\}$ obtained by (\ref{Kp1}) (\ref{Ki}) enable the system to converge to the steady state
($R_1=2$) within 30 iterations, i.e., 0.3s. In contrast, the conservative PI parameters $\{K_{P,l}/5,K_{I,l}/5\}$ take around 120 iterations (i.e., 1.2s)
for the system to converge, while the aggressive PI parameters $\{5K_{P,l},5K_{I,l}\}$ render the system unstable. Similar results are observed for more
complicated topologies, e.g., the 50 users case in Fig. \ref{Symmetric50} that will be introduced in Section \ref{50UsersCase}. Therefore, the PI
parameters given by (\ref{Kp1}) (\ref{Ki}) indeed guarantee system stability with fast convergence.

\subsection{Steady State, Optimality and Fairness}
\begin{figure}[t]
\centering
\includegraphics[width=0.8\linewidth,  trim=40 0 0 0,clip]{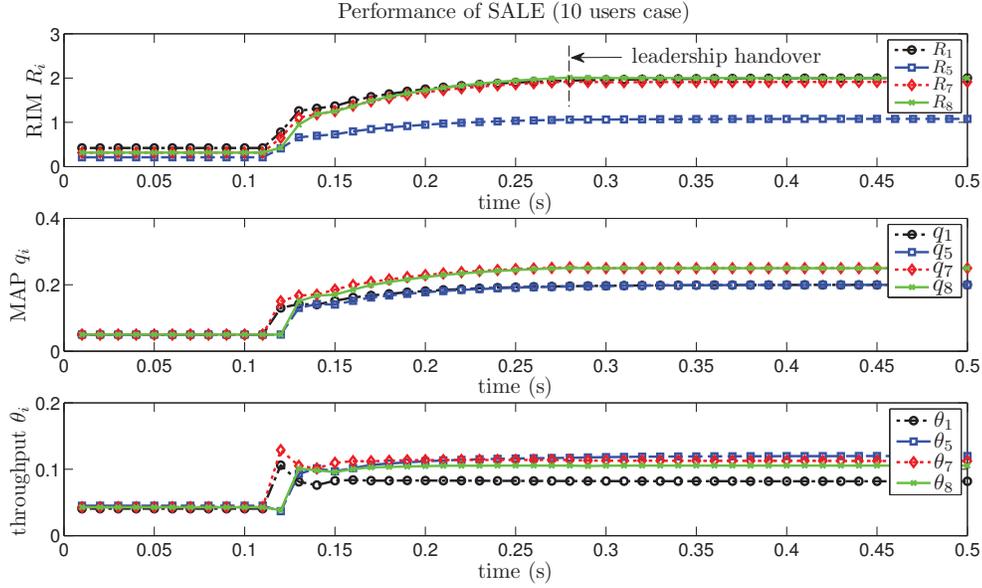} %,height=0.5\linewidth
%\vspace{-0.5em}
\caption{Performance of SALE, Topology in Fig. \ref{Pair10_8}} \label{SALLE157}
%\vspace{-1.5em}
\end{figure}

For the same example given in Fig. \ref{Pair10_8}, we demonstrate more details about the steady state, throughput optimality and fairness among the users.
The transient behaviors of RIM $R_i$, MAP $q_i$ and throughput $\theta_i$ are plotted in Fig. \ref{SALLE157} for users 1, 5, 7 and 8, respectively.

It takes around 40 iterations (i.e., 0.4s) for the system to converge. The preliminary LL election can be completed in the first 10 iterations, in which
each user counts its own ND and exchanges information with its neighbors. Starting from the 11-th iteration, the PI controllers in LL 1 and LL 7 start
working. At the 28-th iteration, user 8 (green line in Fig. \ref{SALLE157}) detects $R_8>2$ and takes over the leadership from the preliminary LL 7 (red
diamond in Fig. \ref{SALLE157}). In the steady state, both LL 1 and LL 8 have $R_1=R_8=2$; the follower 5 has $R_5=1.08$ and has MAP
$\tilde{q}_5=\tilde{q}_1=0.2$; the follower 7 has $R_7=1.91$ and MAP $\tilde{q}_7=\tilde{q}_8=0.25$. Regarding the throughput, generally the users with a
higher ND would have a lower throughput due to contentions from more neighbors. In this example, $\theta_1<\theta_8<\theta_7<\theta_5$.

The distance to Pareto front $d_{Pareto}$ is used to evaluate the throughput optimality. By applying our SALE scheme to Fig. \ref{Pair10_8},
$d_{Pareto}=1.02$, which suggests only 2\% loss between the achieved throughput $\underline{\theta}$ and the Pareto front. Therefore, although the
sub-optimal condition (\ref{LISA}) is used in our design, the result is very close to the actual optimal.

We also evaluate the throughput fairness among the users. When spatial reuse is considered, different users at different spatial locations usually have
different connectivity. As a result, those users with a higher ND usually receive more interference and consequently a lower throughput than those with a
lower ND. Therefore, it is difficult to give an exact measurement of fairness in such a heterogeneous network. We make an attempt to take this spatial
characteristic into consideration, and weigh each user $i$'s throughput by $N_i+1$ (including user $i$ and its neighbors), i.e., we define the
\textit{weighted throughput} for each user $i$ as:
\begin{equation}\label{WeightedThroughput}
\tilde{\theta_i} =(N_i+1)\cdot\theta_i, \forall i \in \mathcal{N}.
\end{equation} %
Then we compute the Jain's fairness index\cite{Jain} for the weighted throughput $\underline{\tilde{\theta}}$:
\begin{equation}\label{JainIndex}
Jain(\underline{\tilde{\theta}})=\frac{(\sum_{i=1}^{N} \tilde{\theta_i})^2}{N\cdot \sum_{i=1}^N \tilde{\theta_i}^2}.
\end{equation}%
Jain's index rates the fairness of an array of values $\underline{\tilde{\theta}}=[\tilde{\theta}_1, \tilde{\theta}_2, \cdots, \tilde{\theta}_N]$ where
there are \textit{N} users and $\tilde{\theta_i}$ is the weighted throughput for the $i$th user. The result ranges from $1/N$ (worst case) to 1 (best
case), and it is maximum when all users receive the same allocation. For our SALE scheme applied to Fig. \ref{Pair10_8},
$Jain(\underline{\tilde{\theta}})=0.9921$, which is close to 1 and suggests good fairness among the users.

%\vspace{-1em}
\subsection{Comparison with Heuristic Algorithm}
Here we apply the heuristic algorithm in \cite{PIMRC} to the same example in Fig. \ref{Pair10_8} and compare performance with that shown in Fig.
\ref{SALLE157}. The algorithm in \cite{PIMRC} assumes no information exchange among the users, and the users adapt their MAPs based on the measured
throughput and channel idle rate, which require a relatively accurate estimation. In the simulations we choose each estimation period to consist of
$L_I=1000$ slots so that the adaptation in the heuristic algorithm works properly. Since the slot time is 0.1ms, each estimation period lasts 0.1s.

\begin{figure}[t]
\centering
\includegraphics[width=0.6\linewidth,  trim=0 0 0 0,clip]{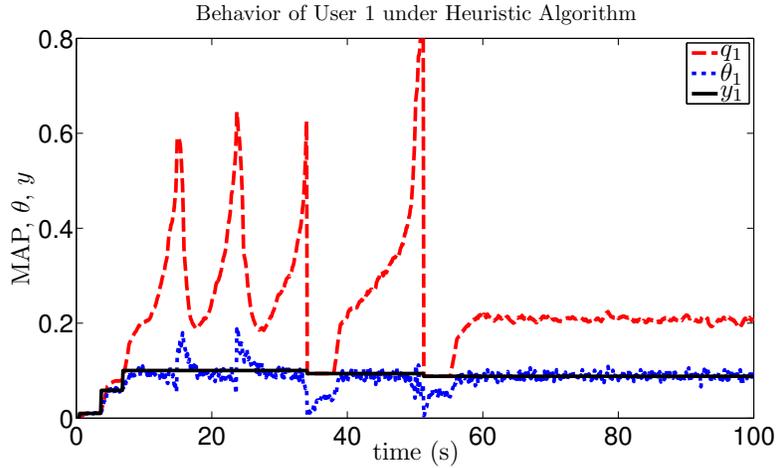} %,height=0.5\linewidth
%\vspace{-0.5em}
\caption{Performance of the Heuristic Algorithm, Topology in Fig. \ref{Pair10_8} } \label{PIMRC10}
%\vspace{-1.5em}
\end{figure}

The transient states of the heuristic algorithm for user 1 are plotted in Fig. \ref{PIMRC10}. As the users heuristically search for the Pareto front, the
system experiences several fluctuations before settling down. The convergence time $T_{conv}$ takes around 600 estimation periods, i.e., around 60s, which
is much longer than that when SALE is used. Since other users experience similar transient states as user 1, their behaviors are not plotted for brevity.
The steady-state performance is summarized in the upper part of TABLE \ref{TableCompare10} for the SALE scheme and the heuristic algorithm. Both schemes
achieve a throughput $\underline{\theta}$ close to the Pareto front ($d_{Pareto}=1.02, 1.045$ respectively), while SALE provides better fairness for the
users (the heuristic algorithm has a lower Jain's index = 0.9685). Since SALE has an additional information exchange overhead of $25/L_S=25/2000=0.0125$,
we also compare the average net throughput $\overline{\theta}_{net,i}$ in the table. Note that we assume both approaches use the same header except the
three additional subfields in the SALE scheme, and the common parts of the header are included in computing the net throughputs.

\begin{table}[t]\footnotesize
\centering
\caption{Comparison between SALE and Heuristic Algorithm}
%\vspace{-0.5em}
\addtolength{\tabcolsep}{-4pt}
\renewcommand{\arraystretch}{1.2}
\begin{tabular}{|c|c|c|c|c|c|c|c|}
\hline
Topology &Scheme &$\overline{\theta_i}$& $\Sigma \theta_i $ & $\overline{\theta}_{net,i}$ &$Jain(\underline{\tilde{\theta}})$& $d_{Pareto}$& $T_{conv}$ \\
\hline
\multirow{2}{*}{Fig. \ref{Pair10_8}} &SALE&0.1246 &1.246 &0.1230 & 0.9921&1.02& $\sim$0.4s\\
\hhline{~-------}
 &Heuristic&0.1158&1.158& 0.1158 & 0.9685&1.045& $\sim$60s\\
\hline
\multirow{2}{*}{Fig. \ref{Symmetric50}} &SALE&0.0578&2.889& 0.0571 &0.9906&1.025& $\sim$0.4s\\
\hhline{~-------}
&Heuristic&0.0535&2.673& 0.0535 & 0.9593&1.04& $\sim$90s\\
\hline
\end{tabular}
%\vspace{3pt}
 \label{TableCompare10}
 %\vspace{-1em}
\end{table}

\subsection{Scalability of SALE}\label{SimulationScalability}

\subsubsection{50 Users Case}\label{50UsersCase}
\begin{figure}[t]
\centering
\includegraphics[width=0.45\linewidth,  trim=40 30 40 30,clip]{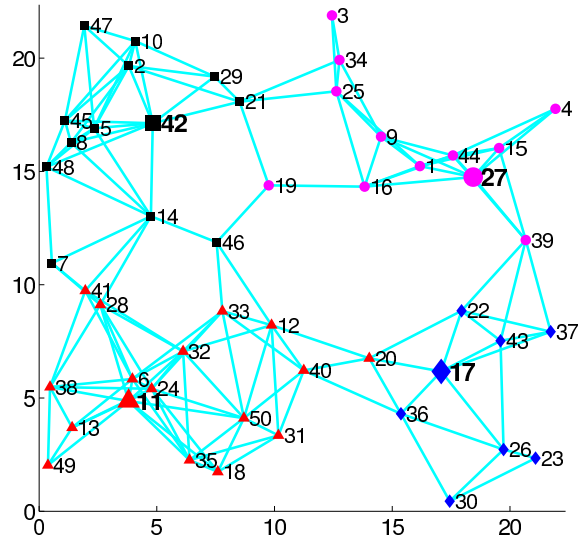} %,height=0.5\linewidth
%\vspace{-0.5em}
\caption{Randomly Generated Connected Topology with 50 Users} \label{Symmetric50}
%\vspace{-1.5em}
\end{figure}

Consider a distributed network with $N$ users, which are randomly placed in a square region of a given area. For simplicity, we assume that all the
distances between any transmitter and its designated receiver are much smaller than the distances between any two transmitters, so that a Tx-Rx pair (user)
can be represented by a single node in the topology. We further assume that all users have transmission range of 5 unit length, and those users who are in
each other's transmission range will have significant interference on each other, and the two users are said to be connected. Based on the above
assumptions, we can generate a random connected topology with 50 users in a square region of area 500 (units), as plotted in Fig. \ref{Symmetric50}.

We apply SALE scheme to Fig. \ref{Symmetric50}, and the whole network is shown to be grouped into 4 trees, governed by LLs $l\in \Omega=\{ 11,17,27,42 \}$
respectively. The users in different trees are marked by different shapes and colors, e.g., the largest tree is marked with red triangles, which is
governed by LL 11 with the highest ND $N_{11}=10$. For comparison, we also apply the
heuristic algorithm to the example in Fig. \ref{Symmetric50}. The performance of the two schemes is summarized in the lower part of TABLE
\ref{TableCompare10}.

The heuristic algorithm converges in around 90s, longer than the 10 users case in Fig. \ref{Pair10_8}. Therefore, the convergence time increases with the
network size in the heuristic algorithm. In contrast, SALE still converges in around 0.4s, thus is more efficient as the network size increases. After the
system converges, both schemes achieve throughputs $\underline{\theta}$ close to the Pareto front ($d_{Pareto}=1.025, 1.04$ respectively), while SALE
provides better fairness for the users, with a higher Jain's index $0.9906>0.9593$.

\subsubsection{100 Users with Various User Density}
We define the \textit{user density} (UD) as the number of users per unit area. For the 50 users case above, the UD is $50/500=0.1$. Now we generate a
sequence of random connected topologies with various UD, by randomly scattering 100 users in a square region of various areas. In particular, when the
square region has an area of 12.5, the network would become fully connected (the diagonal line length is equal to the transmission range of 5). We apply
SALE scheme to these topologies and summarize the results in the upper part of TABLE \ref{ScalabilityTable}.

\begin{table}[t]\footnotesize
\centering
\caption{Scalability of the SALE Scheme}
\addtolength{\tabcolsep}{-4pt}
\begin{tabular}{|c|c|c|c|c|c|c|c|c|}
\hline
Users & Area & UD &$\Sigma \theta_i$ &$Jain(\underline{\tilde{\theta}})$&$d_{Pareto}$& $T_{conv}$& Leaders & $\max H_l$\\
\hline
\multirow{7}{*}{100}& 12.5&8&0.370&1.0000&1& $\sim$0.8s& 1&1\\
\hhline{~--------}
&31.25 & 3.2&0.407&0.9998&1.00& $\sim$0.8s&1&2\\
\hhline{~--------}
&62.5 & 1.6&0.524&0.9912&1.03& $\sim$0.8s& 1&2\\
\hhline{~--------}
&125 &0.8 &0.925&0.9881&1.05& $\sim$0.4s& 1&2\\
\hhline{~--------}
& 250& 0.4&1.492&0.9795&1.055& $\sim$0.4s& 1&4\\
\hhline{~--------}
&500 &0.2 &2.515&0.9856&1.04& $\sim$0.4s& 2&4\\
\hhline{~--------}
&1000 &0.1 &5.193&0.9823&1.02& $\sim$0.4s& 5&5\\
\hline
200& 2000&\multirow{5}{*}{0.1} &9.950&0.9800&1.01& $\sim$0.4s& 12&5\\
\hhline{--~------}
400& 4000& &18.88&0.9692&1.01& $\sim$0.4s&22&8\\
\hhline{--~------}
600& 6000& &27.26&0.9757&1.02& $\sim$0.4s& 32&5\\
\hhline{--~------}
800& 8000& &37.22&0.9771&1.015& $\sim$0.4s& 49&5\\
\hhline{--~------}
1000& 10000& &45.73&0.9799&1.01& $\sim$0.4s& 54&5\\
\hline
\end{tabular}
\label{ScalabilityTable}
\end{table}%

As UD increases from 0.1 to 8 (tend to a fully connected network), the network is grouped into fewer but bigger trees, and the maximum height of all trees
gradually decreases to 1. As UD increases, the total throughput decreases due to increased interference from more neighbors experienced by each user.
However, regardless of the UD, the number of iterations for convergence is still around 40 iterations. Such fast convergence is guaranteed by the
Ziegler-Nichols rules which adapt the PI parameters in (\ref{Kp1}) (\ref{Ki}) to various UDs (associated with different ND $N_l$ at the LL).

As UD increases above 1.6, the individual throughput drops significantly, and packet collision probability increases and affects the success rate of
passing the subfield information. Hence we choose a larger frame of $L_F=200$ slots in each iteration for the proposed SALE scheme. As a result, each
iteration in these high-density topologies now takes 20ms and the convergence time is around 0.8s. Note that similar problems exist for the heuristic
algorithm in dense topologies, in which the individual user throughput is relatively small. To acquire a relatively accurate estimation for a small
throughput value, a longer estimation period is required to suppress the relative error and keep the variance of the estimated throughput at a low level,
in order for the MAP adaptation in the heuristic algorithm to work properly.

Regardless of the UD, the network still achieves close-to-Pareto-front throughput, with $d_{Pareto}$ around 1.05, i.e., only 5\% below the Pareto front. In
particular, $d_{Pareto}$ is equal to 1 in the fully connected case, thus verifying the statement in Section \ref{SimpleSteady} that our SALE scheme
achieves a throughput on the Pareto front in a fully connected network. Finally, the SALE scheme provides good fairness for all users, with Jain's index
around 0.98 (close to 1).

\subsubsection{200$\sim$1000 Users Cases}
Here we keep the UD as 0.1, and increase the number of users by enlarging the area under consideration. Then we generate a sequence of random connected
topologies with $N=200, 400, 600, 800$ and 1000, respectively. We apply the SALE scheme to these topologies and summarize the results in the lower part of
TABLE \ref{ScalabilityTable}.

As the number of users increases, the number of LLs also increases correspondingly, i.e., the whole network is grouped into more trees. As a result, the
average number of users in a tree, as well as the tree height, will not change significantly as the network size grows. This is verified in TABLE
\ref{ScalabilityTable} that the maximum tree height $H_l$ remains around 6 regardless of the network size. Therefore, when a LL $l$ reaches the steady
state ($R_l=2$), the corresponding MAP $q_l$ would only take around 6 hops to reach all the followers in tree $l$. The SALE scheme converges in around 40
iterations, i.e., 0.4s, regardless of the network size, hence providing fast convergence with high scalability.

In the steady state, SALE achieves a throughout $\underline{\theta}$ close to the Pareto front, with $d_{Pareto}$ around 1.02 for all cases. Consequently,
the total throughput $\Sigma \theta_i$ increases almost linearly with the number of users in the network. Meanwhile, the Jain's index is around 0.97 for
all cases, suggesting good fairness among the users.

\section{Conclusion and Future Work}\label{Summary}
This paper focuses on spatial Aloha networks, and attempts to approach global optimization of network throughput based on limited spread of local
information and realize the model's quick convergence and stability. Our proposed SALE scheme is introduced, which can be autonomously implemented by users
using only local information. Specifically, a user with maximum ND in a certain neighborhood is elected as the LL, and the remaining users in this
neighborhood simply follow the same MAP. The SALE scheme makes use of a sufficient condition previously derived for the spatial Aloha network, which
ensures the network to operate in the stable region if the RIM parameter $R$ at the LL(s) satisfies $R\leq 2$. In our design, the LL adjusts its MAP by a
build-in PI controller to achieve $R=2$. The resulting control system is sustained by mathematical foundations from control theory, which guarantees fast
convergence and network stability. Most importantly, RIM is a local parameter based on only local information, which makes the SALE scheme easy and
systematic to implement with high scalability. Through simulations, we validate the fast convergence of the system to a steady-state operating point with
close-to-Pareto-front throughputs and good fairness among the users, while comparing with our previous heuristic algorithm. Future work can extend the SALE
scheme to scenarios with dynamic topology changes\cite{JMchenDataGatheringTNET}, and with multiple channels\cite{JMchenMultiChannel2,LyuMLSG}.

\bibliography{IEEEabrv,AlohaGames}

\end{document}